\documentclass[11pt]{article}

\usepackage{latexsym,bm,graphicx,color,xcolor,nicefrac,titletoc,enumerate,amsmath,amssymb,xfrac,xcolor,physics,cite,setspace}

\usepackage{mlmodern}
\usepackage[T1]{fontenc}

\usepackage[nottoc]{tocbibind} 

\usepackage{hyperref}
\hypersetup{linktocpage=true,colorlinks=true,linkcolor=blue,citecolor=blue,urlcolor=blue}

\usepackage[skip=3pt plus1pt, indent=20pt]{parskip}

\usepackage[bottom]{footmisc}

\usepackage{geometry}
\usepackage{titlesec}
\titleformat{\section}{\large\bfseries\sffamily}{\thesection}{0.5em}{}
\titleformat{\subsection}{\normalfont\bfseries\sffamily}{\thesubsection}{0.5em}{}
\titleformat{\subsubsection}{\normalsize\itshape\sffamily}{\thesubsubsection}{0.5em}{}
\titleformat*{\paragraph}{\normalsize\bfseries\sffamily}

\numberwithin{equation}{section}

\def\a{\alpha}
\def\b{\beta}

\def\G{\Gamma}

\def\k{\kappa}
\def\l{\lambda}
\def\L{\Lambda}
\def\m{\mu}
\def\n{\nu}
\def\r{\rho}
\def\s{\sigma}

\def\o{\omega}

\def\pd{\partial}

\def\pr{\prime}

\def\qq{\quad\quad}
\def\q{\quad}

\newcommand{\cG}{\mathcal{G}}

\newcommand{\cL}{\mathcal{L}}

\newcommand{\cO}{\mathcal{O}}

\newcommand{\Ired}{I_{\text{red}}}
\newcommand{\Lred}{L_{\text{red}}}

\newcommand{\mail}[1]{\href{mailto:#1}{{\tt #1}}}

\begin{document}
	
		
	\begin{center}
		{\Large \bf \sffamily Regular AdS\textsubscript{3} black holes from a regularized Gauss-Bonnet coupling}
	\end{center}
		
	\begin{center}
		\vspace{10pt}
			
		{{\bf \sffamily G{\"o}khan Alka\c{c},}${}^{a}\,${\bf \sffamily  Murat Mesta}${}^{b}\,$ {\bf \sffamily and G{\"o}n\"{u}l \"{U}nal}${}^{c}$}
		\\[4mm]
			
		{\small 
		{\it ${}^a$Department of Aerospace Engineering, Faculty of Engineering,\\ At{\i}l{\i}m University, 06836 Ankara, T\"{u}rkiye}\\[2mm]
				
		{\it ${}^b$Department of Electrical and Electronics Engineering, Faculty of Engineering,\\ At{\i}l{\i}m University, 06836 Ankara, T\"{u}rkiye}\\[2mm]
		
		{\it ${}^c$Department of Biomedical Engineering, Faculty of Engineering,\\ Başkent University, 06790 Ankara, T\"{u}rkiye}\\[2mm]
				
		{\it E-mail:} {\mail{alkac@mail.com}, \mail{murat.mesta@atilim.edu.tr}, \mail{gunalmesta@baskent.edu.tr}}
		}
		\vspace{2mm}
		\end{center}
		
		\centerline{{\bf \sffamily Abstract}}
		\vspace*{1mm}
		\noindent We obtain a three-dimensional bi-vector-tensor theory of the generalized Proca class by regularizing the Gauss-Bonnet invariant within the Weyl geometry. We  show that the theory admits a regular AdS$_3$ black hole solution with primary hairs. Introducing a deformation in the theory, a different regular AdS$_3$ black hole solution is obtained. Charged generalizations of these solutions are given by coupling to Born-Infeld electrodynamics.
		\par\noindent\rule{\textwidth}{0.5pt}
		\tableofcontents
		\par\noindent\rule{\textwidth}{0.5pt}

\section{Introduction}\label{Intro}
Thanks to its relative simplicity, gravity in three dimensions (3d) serves as a useful theoretical laboratory. For example, 3d general relativity (GR) with a negative cosmological constant admits a black hole solution named as the Banados--Teitelboim--Zanelli (BTZ) black hole \cite{Banados:1992wn,Banados1993}, which is locally isomorphic to the AdS$_3$ spacetime but still has all the important properties of a black hole. The geometry is characterized by the mass and the angular momentum, making it globally different than the AdS$_3$ spacetime.  It also has a well-defined event horizon and laws of black hole thermodynamics are satisfied.  One of the most important challenges of a quantum theory of gravity is to provide a microscopic derivation of the black hole entropy. One motivation to study gravity in 3d is that it can be derived without any direct reference to string theory or supersymmetry \cite{Strominger:1997eq} by employing AdS$_3$/CFT$_2$  correspondence (see \cite{Kraus:2006wn} for a review). 

 Another important problem in gravitational physics is that there is no way to avoid singularities at the center of black holes or in the early universe within the standard framework of general relativity \cite{Penrose:1964wq, Hawking:1967ju, Hawking:1970zqf}. This leads to the divergence of certain physical observables such as tidal forces, and therefore looking for some mechanisms solving the problem within GR or trying to modify the theory such that the resulting solutions have no singularities are needed. In the light of the success achieved in the microscopic origin of the black hole entropy, one might try to study this singularity problem in 3d to gain insight into the problem.

The most natural approach to deal with the singularity problem is to try to obtain a regular black hole solution of GR by assuming a perfect fluid as a source (see \cite{Estrada:2020tbz} for such an investigation in 3d).  Although it is possible to calculate the global charges and study the thermodynamics of solutions obtained with this procedure, there is not much information available about a potential mechanism or a modification of GR that would resolve the singularities. Another possible method is coupling GR to non-linear electrodynamics and successful examples can be found in \cite{Cataldo:2000ns, He:2017ujy, HabibMazharimousavi:2011gh}. 

Taking an effective field theory perspective, GR should only be seen as the leading term of an infinite series where the higher order terms are probed at high energies. Therefore, it is particularly important to obtain regular black hole solutions from theories with higher curvature corrections. The first 3d example was obtained in \cite{Bueno2021} (see \cite{Bueno:2025dqk} for the charged generalization), where a particular scalar-tensor theory, termed as the generalized electromagnetic quasi-topological gravity, admits regular black hole solutions provided that the scalar field is chosen to be linear in the azimuthal angle.

While the most natural candidates for admitting exact black hole solutions are higher-order pure gravity theories with second-order field equations for the metric, the most general theory with this property, the Lovelock gravity \cite{Lovelock:1971yv,Lovelock:1972vz,Padmanabhan:2013xyr}, is only defined in $d>4$. The reason is that the lowest order correction to the Einstein-Hilbert action, the Gauss-Bonnet (GB) invariant, is a topological term in 4d and identically vanishes in 3d. Recently, it was shown in \cite{Charmousis:2025jpx} that a well-defined coupling that contributes to field equations in 4d can be obtained through a regularization of the GB invariant within the Weyl geometry. 

The regularization is based on the fact that the GB invariant is non-dynamical in $d<5$ independently of the connection that is used. Therefore, evaluating the difference between the GB invariants for two different connections in a generic $d$ dimension, one obtains, up to a boundary term, a non-trivial Lagrangian density multiplied by $(d-3)(d-4)$, which can be removed by scaling the coupling in $d$ dimensions. This way, two different regularized versions of the Gauss-Bonnet coupling can be obtained in 3d and 4d. Considering one Weyl connection and one Christoffel connection, we can obtain a vector-tensor theory of the generalized Proca class \cite{Heisenberg:2014rta}. On the other hand, starting with two Weyl connections leads to a bi-vector-tensor coupling. Having second-order field equations, these vector-tensor theories possess black hole solutions with interesting properties. 

In 4d, these two different possibilities were studied in \cite{Charmousis:2025jpx, Eichhorn:2025pgy}. In the first case, with a single Weyl vector, one finds black hole solutions with a primary hair, i.e. an additional integration constant that modifies the geometry. In the second case, with two different Weyl vectors, regular black hole solutions with two primary hairs arise. The regularity of the resulting solutions is not guaranteed from the form of the Lagrangian density, as, for example, in the case of the regular point charge solution of Born-Infeld electrodynamics. However, the 4d case suggests a natural generalization of these solutions to 3d.

The regularized GB invariant in 3d with a single Weyl vector was constructed in \cite{Alkac:2025zzi} and the resulting black holes with a primary hair was given. Here, we investigate the possibility of having two Weyl vectors and construct a bi-vector-tensor coupling. In complete parallelism with the 4d story, we obtain regular AdS$_3$ black hole solutions. As discussed in the last section, these black hole solutions are expected to exhibit interesting thermodynamical properties and the microscopic derivation of their entropy through the AdS$_3$/CFT$_2$ correspondence is particularly intriguing because of the existence of two primary hairs.

The outline of this paper is as follows: In Section \ref{section:GB}, we discuss the regularization of the GB invariant with one and two Weyl vectors. In Section \ref{section:revisited}, in order to introduce our methodology and notation, we revisit the solution arising from the regularized GB coupling with a single Weyl vector as given in \cite{Alkac:2025zzi}. In Section \ref{section:regularBH}, we present the regularization of the GB invariant with two Weyl vectors. Then, we give regular AdS$_3$ black hole solutions, and charged versions by coupling the theory to Born--Infeld electrodynamics. We present our conclusions in Section \ref{section:conclusion}.

\section{3d regularized Gauss--Bonnet invariant with two Weyl vectors}\label{section:GB}
Lovelock gravity is a higher-curvature generalization of GR that shares many important physical properties of it (see \cite{Padmanabhan:2013xyr} for a review). The lowest order correction that is quadratic in curvature, the GB invariant is given by
\begin{equation}
	 \cG = R^2 - 4 R_{\m\n} R^{\m\n}  + R_{\m\n\r\s} R^{\m\n\r\s}.
\end{equation}
As mentioned in the Introduction, it contributes to field equations only when $d>4$. However, it was realized that there exists regularization mechanisms that allow us to obtain well defined Lagrangians giving rise to a nontrivial dynamical contribution in the lower dimensions. The first example yields scalar-tensor theories of Hordenski class \cite{Horndeski:1974wa,Deffayet:2009mn} and the resulting theories have quite interesting properties (see \cite{Fernandes:2020nbq,Lu:2020iav,Kobayashi:2020wqy,Hennigar:2020lsl,Ma:2020ufk,Hennigar:2020fkv,Hennigar:2020drx,Fernandes:2021dsb,Khodabakhshi:2022knu,Mao:2022zrf,Alkac:2022fuc,Alkac:2022zda,Bakopoulos:2022gdv,Fernandes:2022zrq,Guajardo:2023uix, Alkac:2023mvr,Babichev:2024krm,Fernandes:2025fnz,Fernandes:2025eoc}). Recently, another regularization scheme within the Weyl geometry generating vector-tensor theories of the generalized Proca class was discovered \cite{Heisenberg:2014rta}. Starting from a torsionless Weyl correction $\widetilde{\G}^\a_{\,\m\n}$ with the corresponding covariant derivative $\widetilde{\nabla}_\m$, the Weyl vector $W$ is defined by the following relation
\begin{equation}
   \widetilde{\nabla}_\a g_{\m\n}=-2 g_{\m\n}W_\a,
\end{equation}
from which the Weyl connection can easily be found in terms of the Levi--Civita connection and the Weyl vector (see reference \cite{Dengiz:2011ig, Tanhayi:2012nn, BeltranJimenez:2014iie,Bahamonde:2025qtc} for details). When the GB term is calculated with the Weyl connection, it reads
\begin{equation}
	\widetilde{\cG}[W]=\mathcal{G}+(d-3) \nabla_\mu J^\mu+(d-3)(d-4) \cL,
\end{equation}
where
\begin{align}
	J^\mu&=8 G^{\mu \nu} W_\nu+4(d-2)\left[W^\mu\left(W^2+\nabla_\nu W^\nu\right)-W_\nu \nabla^\nu W^\mu\right], \\
	\cL&=4 G^{\mu \nu} W_\mu W_\nu+(d-2)\left[4 W^2 \nabla_\mu W^\mu+(d-1) W^4\right],
\end{align}
with $W^2=W_\m W^\m$, $W^4=(W^2)^2$, and $\cG$ is the GB invariant calculated with the Levi--Civita connection. Disregarding the boundary terms, the regularized GB invariant with a single Weyl vector is defined as
\begin{equation}
	\cG[W] = \lim_{d \to p} \frac{\widetilde{\cG}[W] - \cG}{d-p}, \qq p = 3,4.
\end{equation}
The two GB terms $\widetilde{\cG}[W]$ and $\cG$, and therefore their difference, do not contribute to the field equations. However, dividing the difference by $d-p$, which corresponds to scaling the relevant coupling constant, and then setting $d=p$ give a Lagrangian which yields non-trivial dynamics in $d=p$. One can repeat the same exercise by taking the difference between two GB invariants with different Weyl vectors as follows \cite{Eichhorn:2025pgy}
\begin{equation}
	\lim_{d \to p} \frac{\widetilde{\cG}[B]- \widetilde{\cG}[C]}{d-p}=\cG[B]-\cG[C].
\end{equation}
As a result, one obtains the difference of the regularized GB invariant for two different Weyl vectors. This coupling, when supplemented by the Einstein-Hilbert term, was shown to admit regular black holes in 4d. Here, we will use the 3d version of this coupling to obtain regular AdS$_3$ black holes. It reads \cite{Alkac:2025zzi}
\begin{equation}\label{GWdef}
\cG[W]=-4 G^{\mu\nu}W_\mu W_\nu-4 W^2 \nabla_\mu W^\mu-2 W^4,
\end{equation} 
for a Weyl vector $W$. 

Before we attempt to find the black hole solutions of the resulting by bi-vector-tensor theory, as a warm-up exercise, we will revisit the solution of the theory with the single Weyl vector given in \cite{Alkac:2025zzi} and explain the shortcut used in \cite{Liu:2025dqg} recently.

\section{Solution in [20] revisited}\label{section:revisited}
We define the regularized vector-tensor Einstein--Gauss--Bonnet theory by the following action
\begin{equation}
    I=\int \dd^3x \sqrt{-g}(R - 2\L_0 + \b\,\cG[W]),
\end{equation}
where $\L_0$ is the bare cosmological constant, $\cG[W]$ is the regularized GB invariant with the Weyl vector $W$ given in \eqref{GWdef} and $\b$ is the corresponding coupling constant. By assuming a line element of the form 
\begin{equation}\label{LineEle}
\dd s^2 = -N^2(r)f(r) \dd t^2+\frac{\dd r^2}{f(r)}+r^2\dd\theta^2
\end{equation}
and the following vector components
\begin{equation}\label{Weqw0plusw1}
   W_\m\dd{x}^\m=w_0(r)\dd{t}+w_1(r)\dd{r},
\end{equation}
one can obtain a reduced action $\Ired = \int \dd{r} L_\text{red}$ where the reduced Lagrangian $\Lred$ is given by
\begin{align}
  L_\text{red}&=L^\text{EH}_\text{red}+\b L_\text{red}^{[W]},
  \\ L^\text{EH}_\text{red} &= N(\L_0r+f^\pr/2),\\
    L_\text{red}^{[W]} &= \frac{r w_0^4}{f^2N^3}
- N^\pr\left[\frac{2r w_0^2w_1}{N^2} + 2 f^2w_1^2(1+rw_1)\right]\nonumber
- \frac{w_0^2 \left[(1+2rw_1)f^\pr+2f(w_1+rw_1^2+rw_1^\pr)\right]}{fN}\nonumber\\ 
&+N\left[fw_1^2(1+2rw_1)f^\pr + f^2w_1^2(2w_1+rw_1^2+2rw_1^\pr)\right].\label{LredW}
\end{align}
Here, $L^\text{EH}_\text{red}$ is the contribution of the Einstein-Hilbert term (with the bare cosmological constant $\L_0$) and $L_\text{red}^{[W]}$ is the contribution of the regularized GB invariant with the Weyl vector $W$. Equations of motion for the functions $(N,f,w_0,w_1)$ can be found by extremizing the action. However, as first observed in \cite{Charmousis:2025jpx} and applied to the 3d theory in \cite{Alkac:2025zzi}, in order to decouple the equations, it is necessary to realize that the equations are invariant under the global symmetry 
\begin{equation}\label{Transformations}
w_0^2\to w_0^2+ 2\l f N (1 + r w_1)+ \l^2 r^2, \qq w_1\to w_1+ \l \frac{r}{N f},
\end{equation}
where $\l$ is a constant parameter. The corresponding Noether charge is as follows
\begin{equation}\label{NoetherCharge}
    Q = \frac{r^2 w_0^2}{f N^2}-(1+r w_1)^2 f.
\end{equation}
Using this Noether charge together with the consistency of equations of motion, one finds that $N^\prime=0$ and therefore, one may take $N=1$. The metric function is found as
\begin{equation}
f_\pm=-(m+q) + \frac{r^2}{2\b}\left(1 \pm \sqrt{1+ 4 \b \L_0 + \frac{4\b q}{r^2}}\,\right).
\end{equation}
We see that an additional integration constant $q$ modifies the geometry, i.e., we have a black hole solution with a primary hair. The vector components can be found in terms of the metric function $f$, and integration constants $(m,q,c)$, where $c$ is another integration constant that does not modify the geometry. The norm of the Weyl vector is related to this constant on-shell, i.e. $W^2=-2c$. It turns out that the global symmetry \eqref{Transformations} with $\l=-c$ kills the effect of the constant $c$ on the vector components, and one obtains a solution with a Weyl vector whose norm vanishes on-shell ($W^2=0$).

The solutions of the theory in general $d$-dimensions are studied in \cite{Liu:2025dqg}. It was observed that, since the ultimate aim is to find the metric function and the norm of the vector field can always be made null by the global symmetry \eqref{Transformations}, one can use the following steps to find the solution in a theory containing the vector-tensor coupling in \eqref{GWdef}:
\begin{enumerate}
    \item  Find the reduced action by using the ansatz (\ref{LineEle}, \ref{Weqw0plusw1}) and derive the equations of motion.
    \item Take
    \begin{equation}\label{w1tow0}
     w_1=\frac{w_0}{Nf},
\end{equation}
with which the norm of the Weyl vector vanishes ($W^2=0$). Thanks to this choice, one can easily get the solution where there is no additional integration constant (other than the hair parameter) appear in the vector components. This is equivalent to setting the additional constant to zero at the end by using the global symmetry of the reduced action, as explained before.
	\item  Then, one sees that the consistency of equations of motion implies $N=1$ and they now become decoupled. Solving $w_0$ first, the metric function $f$ can be easily found.
\end{enumerate}

In the next section, we will apply this procedure to the bi-vector-tensor theory and its deformed version.

\section{Regular black hole solutions}\label{section:regularBH}
Now, we are prepared to study the solutions of the theory defined by the action
\begin{equation}\label{Action}
	I = \frac{1}{16 \pi G} \int \dd^3{x} \sqrt{-g} \left[R - 2 \L_0 + \ell^2 \left( \cG [B]-\cG [C]\right)\right],
\end{equation}
where $\ell$ is a constant of length dimension. For the Weyl vectors $B$ and $C$, similar to \eqref{Weqw0plusw1}, we take
\begin{equation}
    B=b_0(r) \dd t+b_1(r) \dd r,\quad \quad C= c_0(r) \dd t+ c_1(r)  \dd r.
\end{equation}
 The contributions $L_\text{red}^{[B]}$ and $L_\text{red}^{[C]}$ to the reduced Lagrangian can be found by the replacement $W\rightarrow B,C$ in \eqref{LredW}.
 
 In 4d, by introducing a deformation, a different black hole solution was found \cite{Eichhorn:2025pgy}. Therefore, we do the same and consider also the following action with the deformation parameter $\k$
 \begin{equation}\label{Actiondouble}
	I = \frac{1}{16 \pi G} \int \dd^3{x} \sqrt{-g} \left[R - 2 \L_0 + \ell^2  \,\left( \cG [B]-\k\,\cG [C]\right)\right],
\end{equation}
for which the reduced Lagrangian is given by 
\begin{equation}\label{LredDeformed}
    L_\text{red}=L^\text{EH}_\text{red}+\ell^2 \left(L_\text{red}^{[B]}-\k L_\text{red}^{[C]}\right).
\end{equation}
Note that this deformed coupling follows from the regularization

\begin{equation}
	\lim_{d\to3}\frac{\widetilde{\cG}[B]-\k\widetilde{\cG}[C]+(\k-1) \cG}{d-3}=\cG [B]-\k\,\cG [C].
\end{equation}

By the method of \cite{Liu:2025dqg} that we have outlined in the previous section, studying the solutions of the undeformed ($\k=1$) and the deformed ($\k\neq1$) theory is a straightforward exercise.
The equations of motion can be obtained by extremizing the reduced action defined by the Lagrangian in \eqref{LredDeformed}. In order to present them in a more transparent manner, it is useful to consider first the contribution of the regularized Gauss--Bonnet invariant with a Weyl vector $W$ separately. From the reduced action $I_\text{red}^{[W]}=\int\dd{r}\Lred^{[W]}$, where $\Lred^{[W]}$ is given in \eqref{LredW}, we find
\begin{align}
    \Xi[W]=&\,6 r\omega_0^4-2 f N^2\left[\omega_0^2(1+2r\omega_1)-f^2N^2w_1^2(3+2r\omega_1)\right]f^\prime\nonumber
    \\ &+2 f^2N^2\left[f^2N^2\left(-r\omega_1^4+4\omega_1\omega_1^\prime+4r\omega_1^2\omega_1^\prime\right)-2\left(2r\omega_1\omega_0^\prime\omega_0+\omega_0^2(2\omega_1+r\omega_1^2+2r\omega_1^\prime)\right)\right]
    \\\Gamma[W]=&\,4r \omega_0^4-2f^4N^3\omega_1^2(3+2r\omega_1)N^\prime-4f^4N^4\omega_1(1+r\omega_1)(\omega_1^2-\omega_1^\prime)\nonumber
    \\ &-4f^2N^2\omega_0(1+2r\omega_1)\omega_0^\prime+2f^2N\omega_0^2\left((1+2r\omega_1)N^\prime-2N(\omega_1+r\omega_1^\prime)\right)\\
    \Psi[W]=&-2r \omega_0^2+f N\left( N\left(1+2r\omega_1\right)f^\prime+2 f\left(r\omega_1 N^\prime+N\left(\omega_1+r \omega_1^2+r \omega_1^\prime\right)\right)\right)
    \\\Phi[W]=&-r N\omega_0^2f^\prime+f^2 N^3\omega_1(1+r\omega_1)f^\prime+ 2 f^3 N^2 \omega_1(1+2r\omega_1)(N\omega_1+N^\prime)\nonumber
    \\&-2 r f \omega_0 \left(\omega_0 N^\prime+N(\omega_0\omega_1- \omega_0^\prime)\right)
\end{align}
where 
\begin{equation}
    \Xi[W]=\fdv{I_\text{red}^{[W]}}{N},\q \Gamma[W]=\fdv{I_\text{red}^{[W]}}{f},\q \Psi[W]=\fdv{I_\text{red}^{[W]}}{\o_0},\q\Phi[W]=\fdv{I_\text{red}^{[W]}}{\o_1}.
\end{equation}
With these at hand, the equations of motion of the bi-vector theory can be written as
\begin{align}
    & \Psi[B]=0,\qq\Psi[C]=0,\qq\Phi[B]=0,\qq\Phi[C]=0,\label{eom1}\\
    & -f^2 N^4(2r\Lambda_0+f^\prime)+\ell^2(\Xi[B]-\kappa \Xi[C])=0,\qq f^3N^3 N^\prime+\ell^2(\Gamma[B]-\kappa \Gamma[C])=0\label{eom2}
\end{align} 

 We now present regular black hole solutions and their charged versions by coupling the theories to Born--Infeld electrodynamics. In Appendix \ref{appendix}, charged black hole solutions obtained by coupling to Maxwell's electrodynamics are given. Due to the logarithmic form of the Coulomb potential in 3d, they possess a curvature singularity.

\subsection{Undeformed theory ($\k=1$)}
Analogous to \eqref{w1tow0}, we choose
\begin{equation}\label{b1c1}
    b_1=\frac{b_0}{Nf},\qq\qq c_1=\frac{c_0}{Nf}.
\end{equation}
Then, the consistency of the equations (\ref{eom1}, \ref{eom2}) forces $N=1$, and one finds
\begin{equation}\label{b0c0}
    b_0=\frac{q_b}{r}-\frac{f}{2r},\quad\quad c_0=\frac{q_c}{r}-\frac{f}{2r},
\end{equation}
where $q_b$ and $q_c$ are the (constant) hair parameters. By the choice \eqref{b1c1} and the condition $N=1$, the Weyl vectors $B$ and $C$ are null on-shell. With these vector components, the metric function can be found to be
\begin{equation}\label{metric_1}
   f = \frac{r^2(-\L_0r^2 - m)+4\ell^2(q_b^2-q_c^2)}{r^2+4\ell^2(q_b-q_c)},
\end{equation}
where we have introduced another integration constant $m$. As $r\to \infty$, the metric function behaves as
\begin{equation}
    f=-\L_0 r^2- m+ 4 \ell^2 (q_b-q_c)\L_0+\cO(1/r^2),
\end{equation}
which shows that the effective cosmological constant is equal to the bare cosmological constant introduced in the action ($\L=\L_0$). Later, we will see that this is not case for the solutions of the deformed theory.

\begin{figure}
    \centering
    \includegraphics[width=\linewidth]{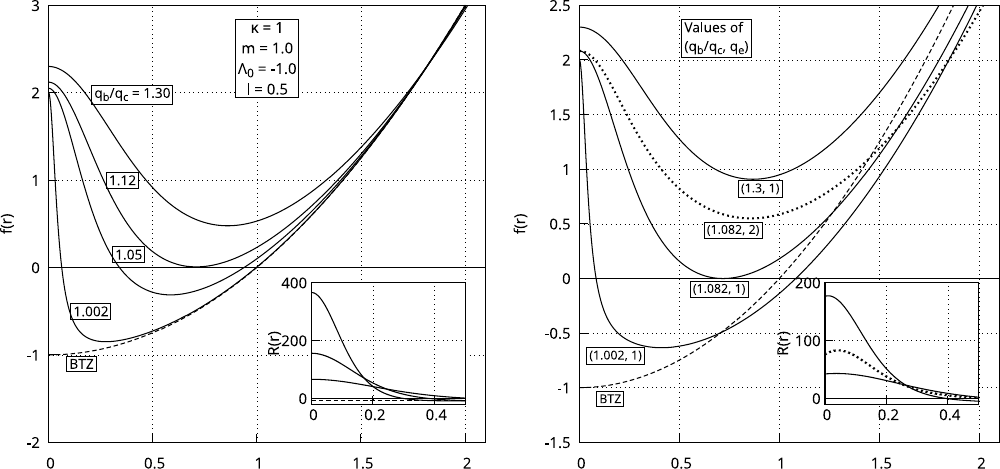}
    \caption{Figures showing the metric function and the corresponding curvature scalars for the undeformed theory. On the left, the plot of the metric function of the uncharged black hole in \eqref{metric_1} is shown for a fixed set of parameters ($\k=1$, $m=1$, $\Lambda_0=-1$, and $\ell=0.5$) and for various values of the ratio $q_b/q_c$ (noted in the text boxes in the figure). The dashed line represents the BTZ black hole solution. On the right, metric function of the charged black hole in \eqref{metric_2} is shown. The fixed parameters are the same with that of plot on the left, and the values of $q_b/q_c$ and $q_c$ varies as shown in the text boxes. The dotted line indicates the effect of a larger electric charge $q_e$ on the metric. In the insets, one can see the regularity of the curvature scalars explicitly. }
    \label{fig1}
\end{figure}

As long as $q_b>q_c$, any singularity at a positive $r$ is avoided. Near the origin $r\to 0$, one finds
\begin{equation}\label{Ricci1}
    R=\frac{3(m+q_b+q_c)}{2\ell^2(q_b-q_c)} + \cO(r^2),
\end{equation} 
and the only other independent curvature invariants in 3d ($R^\mu_{\ \nu} R^\nu_{\ \mu}$ and $R^\mu_{\ \r} R^\r_{\ \nu} R^\n_{\ \mu}$) \cite{Paulos2010,Gurses2012,Bueno2022} are also finite at the origin. For different sets of parameters, this solution is either a regular horizonless spacetime or a regular black hole with a double horizon. For a fixed parameter set, the metric function given in \eqref{metric_1} has the behavior shown in Figure \ref{fig1} (left), where we emphasize the effect of $q_b/q_c$ ratio on the horizon structure. While $q_b/q_c=1.05$ gives an extremal black hole solution, where the horizons coincide, any smaller and larger values $q_b/q_c$ lead double and horizonless cases respectively. 

Note that  $q_b=q_c$ makes $B_\m=C_\m$ on-shell. This is a case where the effective energy-momentum tensor vanishes on-shell and one obtains a stealth BTZ black hole solution, whose first example in the 3d literature was given in \cite{Ayon-Beato:2004nzi}.

\subsection{Undeformed theory ($\k=1$) coupled to Born--Infeld electrodynamics}\label{subsection:k=1BI}
We now couple the theory to Born--Infeld electrodynamics with the Lagrangian \cite{Cataldo:1999wr}
\begin{equation}
   \cL_\text{BI}= b^2 \left( 1 - \sqrt{1 + \frac{F_{\m\n}F^{\m\n}}{2 b^2}}\, \right),
\end{equation}
where $F_{\m\n}=\pd_\m A_\n-\pd_\n A_\m$ is the field strength tensor of the $U(1)$ gauge field $A_\m$. In the $b\to\infty$ limit of the Born-Infeld parameter, it reduces to the Lagrangian of Maxwell's theory in \eqref{LMax}. By assuming
\begin{equation}\label{FieldA}
    A_\m\dd{x}^\m=\phi(r)\dd{t},
\end{equation}
the scalar potential solving the equation of motion can be taken as \cite{Alkac2022a}
\begin{equation}
    \phi = -q_e \log\left(\frac{r + \sqrt{r^2 + \frac{q_e^2}{b^2}}}{2}\right).
\end{equation}
Using the vector components in (\ref{b1c1}, \ref{b0c0}) together with the contribution from the Born--Infeld electrodynamics, the metric function is found as
\begin{equation}\label{metric_2}
    f=\frac{r^2[-\L_0r^2-m+\frac{q_e^2}{2}(\frac{1}{1+\psi}-\text{arctanh}(\psi))]+4\ell^2(q_b^2-q_c^2)(\frac{b^2r^2}{q_e^2}+1)}{r^2+4\ell^2(q_b-q_c)}
\end{equation} 
where \begin{equation}
    \psi = \frac{r}{\sqrt{r^2+q_e^2/b^2}}.
\end{equation}
The effective cosmological constant is again equal to the bare one ($\L=\L_0$). Near the origin, the Ricci scalar behaves as
\begin{equation}\label{Ricci2}
    R=\frac{3[-q_e^4+2q_e^2(q_b+q_c+m)-8b^2\ell^2(q_b^2-q_c^2)]}{4\ell^2q_e^2(q_b-q_c)}+ \cO(r^2)
\end{equation}
and the other curvature invariants are well-defined at the origin, which show that it is a regular black hole as long as $q_b>q_c$. In Figure \ref{fig1} (right), the behavior of the metric function in \eqref{metric_2} for different values of the electric charge $q_e$ can be seen. Considering the same fixed parameter set of parameters with that of the uncharged solution, presence of the electric charge shifts the critical $q_b/q_c$ ratio to larger values. For example, $q_e=1$ leads to an extremal black hole for $q_b/q_c=1.08$ instead of $1.05$. Increasing $q_e$ to 2, the metric function typically has larger values for all $r$.

When $q_b=q_c$, we have the black hole solution of Einstein--Born--Infeld theory \cite{Cataldo:1999wr} as a stealth solution.

\subsection{Deformed theory ($\k\neq1$)}
The deformed theory can also be solved through the same procedure and the vector components take the same form in (\ref{b1c1}, \ref{b0c0}). While a general solution for the metric function is possible to obtain, we prefer not to present it here due to its cumbersome form. As in 4d \cite{Eichhorn:2025pgy}, by choosing a particular relation between the hair parameters, i.e. $q_c=q_b/\k$, we can find a simpler solution given by
\begin{equation}\label{metric_3}
    f_\pm = -\frac{r^2}{2(\k-1)\ell^2}\left(1 \pm \sqrt{1-4\L_0(\k-1)\ell^2+\frac{4(\k-1)^2\ell^4(4q_b^2-\k mr^2)}{\k r^4}}\,\right).
\end{equation}
The sign in front of the square root can be fixed by demanding a well defined $\ell\to0$ limit. We have
\begin{align}
    f_+&=-\frac{r^2}{(\k-1)\ell^2}+\Lambda_0r^2+\cO(\ell^2),
    \\ f_-&=-\Lambda_0r^2+\cO(\ell^2).
\end{align} which means that $f_-$ branch should be chosen. The effective cosmological constant can be read from the $r\to\infty$ limit as
\begin{equation}\label{cosmo}
    \Lambda=\frac{1-\sqrt{1-4(\k -1)\ell^2\Lambda_0}}{2(\k-1)\ell^2}.
\end{equation}
For an AdS$_3$ black hole ($\L<0$), $\L_0$ should be negative. When $\k>1$, the solution is guaranteed to be asymptotically AdS$_3$. However, when $\k<1$,  the following additional constraint should be satisfied
\begin{equation}
    \L_0\geq\frac{1}{4(\k-1)\ell^2}.
\end{equation}
As $r\to0$,  the Ricci scalar (and other curvature invariants) remain finite.  Therefore, this is again a regular black hole solution.

\begin{figure}
    \centering
    \includegraphics[width=\linewidth]{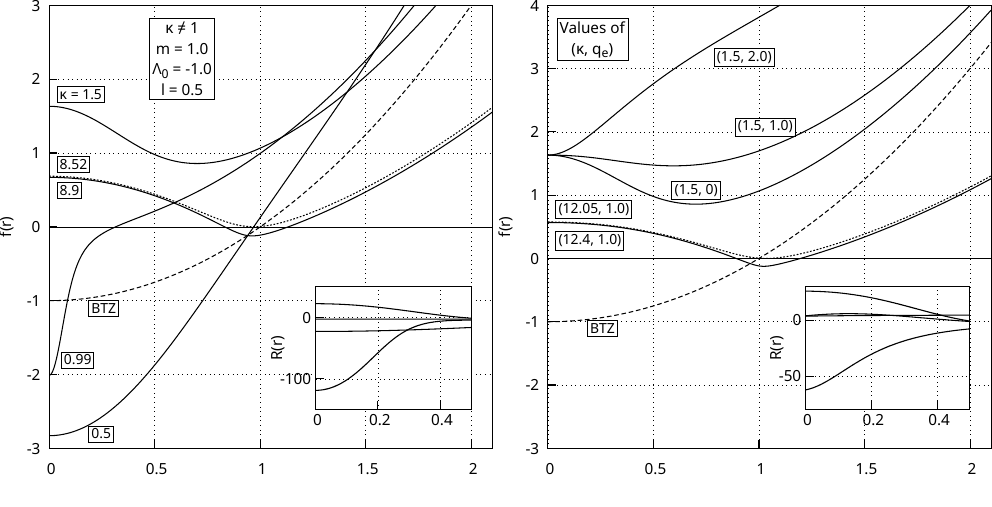}
    \caption{Figures showing the metric function and the corresponding curvature scalars for the deformed theory. The same parameters that are used in Figure \ref{fig1} are implemented here as well as $q_b=1$. On the left, the effect of the deformation parameter $\kappa$ on the metric function \eqref{metric_3} can be seen, giving rise to different horizon structures. The dotted line is for the extremal case corresponding to $\kappa=8.52$. The dashed line represents the BTZ black hole solution. On the right, the metric function \eqref{metric_4} of the charged black hole is shown for different $\kappa$ values, and $q_e=1$ and $q_e=2$. The presence of the electric charge $q_e$ increases the required $\kappa$ value for the extremal solution to 12.05.}
    \label{fig2}
\end{figure}

Unlike the undeformed theory, this solution might also correspond to black holes with a single horizon in addition to horizonless geometries and black holes with double horizon. As an example, we consider a case with $m=1$, $\Lambda_0=-1$, $\ell=0.5$, and $q_b=1$ and study the effect of the deformation parameter $\k$ on the solution. We find the following behavior,
\begin{equation}\label{matrix}
\begin{matrix}
     \k<0:& \text{The solution is not real.}\\
     0<\k<1:& \text{Black hole with a single horizon.}\\
     1<\k<8.52:& \text{Horizonless geometry.}\\
     \k=8.52:& \text{Extremal black hole.}\\
     8.52<\k<9.0:& \text{Black hole with double horizon.}\\
     \k\geq 9.0:& \text{The solution is not real.}\\
\end{matrix}
\end{equation}
In Figure \ref{fig2} (left), these different possibilities together with the regular behavior of the Ricci scalar are demonstrated.

\subsection{Deformed theory ($\k\neq1$) coupled to Born--Infeld electrodynamics}
By the same procedure applied in Subsection \ref{subsection:k=1BI}, the metric function is obtained as
\begin{align}
    f&=-\frac{r^2}{2(\k-1)\ell^2}\times\nonumber\\
    &\left[1-\sqrt{1-\frac{2(\k-1)\ell^2(q_e^2+b^2r^2)(b^2(\psi-1)+2\L_0)}{b^2r^2}-\frac{2(\k-1)\ell^2q_e^2\text{arctanh}(\psi)}{r^2}+\frac{4(\k-1)^2\ell^4(4q_b^2-r^2\k m)}{\k r^4}}\,\right]\label{metric_4}
\end{align}
The effective cosmological constant $\Lambda$ is the same as in \eqref{cosmo}. It is straightforward to show that this is also a regular black hole solution. In Figure \ref{fig2} (right) one can see that the overall behavior of the solution is quite similar to the uncharged case summarized in \eqref{matrix}. A nonzero electric charge $q_e$ shifts the upper and lower values of the deformation parameter $\k$ generating different horizon structures towards larger values.

\section{Conclusion}\label{section:conclusion}
In this paper, following \cite{Fernandes:2025fnz}, we obtained a 3d bi-vector-tensor theory by regularizing the GB invariant with two Weyl vectors and showed that it admits a regular AdS$_3$ black hole solution with two distinct primary hair parameters. By introducing a deformation and choosing a particular relation between hair parameters, we obtain a different regular AdS$_3$ black hole. Coupling the theories to Born-Infeld electrodynamics yields regular charged black holes.

With these solutions, it is possible to ask questions about the black hole interior for a regular black hole in 3d. Previous studies show that regular AdS$_3$ black holes are geodesically complete and tidal forces experienced by observers are finite everywhere \cite{Bueno:2025dqk, Fernandes:2025eoc}. The same behavior is expected for the solutions presented here. Because of the hair parameters, the thermodynamical properties of the black hole solutions given here might be particularly interesting (see \cite{Priyadarshinee:2023cmi, Daripa:2024ksg} for previous 3d examples).

The microscopic derivation of entropy by using the AdS$_3$/CFT$_2$ correspondence was generalized to solutions with scalar hair in \cite{Correa:2010hf, Correa:2011dt, Correa:2012rc}. Instead of the usual form of the Cardy formula for 2d CFTs with the central charge of the theory, an alternative version with the ground state energy was shown to successfully reproduce the semi-classical entropy of black holes with scalar hair. The ground state is identified on the gravitational side with the soliton solution obtained by a double Wick rotation from the static black hole solution. Together with the solutions in \cite{Alkac:2025zzi}, we expect this procedure to work for solutions given here and we hope to report more on this in near future.

\appendix
\section{Coupling to Maxwell's electrodynamics}\label{appendix}
In this appendix, we give the resulting charged black hole solutions when the undeformed ($\k=1$) and the deformed ($\k\neq1$) theory are coupled to Maxwell's electrodynamics with the Lagrangian
\begin{equation}\label{LMax}
    \cL_\text{M}=-\frac{1}{4} F_{\m\n} F^{\m \n}.
\end{equation}
Taking the vector field as in \eqref{FieldA}, one finds the Coulomb potential
\begin{equation}
    \phi=-q_e\log(r).
\end{equation}
The components of the Weyl vectors are given in (\ref{b1c1}, \ref{b0c0}), and the metric functions are as follows
\begin{align}
\k=1:\qq f &= \frac{r^2\left[-\L_0r^2-m-\frac{q_e^2}{2} \log(r)\right]+4\ell^2(q_b^2-q_c^2)}{r^2+4\ell^2(q_b-q_c)},\\
\k\neq1:\qq f &= -\frac{r^2}{2\ell^2(\k-1)}\times\nonumber\\
&\left(1-\sqrt{1-4\L_0(\k-1)\ell^2+\frac{4(\k-1)^2\ell^4(4q_b^2-\k m r^2)}{\k r^4}-\frac{2(\k-1)q_e^2\ell^2\log(r)}{r^2}}\,\right).
\end{align}
Note that the relation between the hair parameters $q_c = \frac{q_b}{\k}$ was used in obtaining the latter solution as done in the main text. From the $r\to0$ limit, the behavior of the Ricci scalar is found as
\begin{align}
    \k=1:\qq R &= \frac{12(q_b+q_c+m)+5q_e^2+6q_e^2\log(r)}{8\ell^2(q_b-q_c)} + \cO(r^2),\\
    \k\neq1:\qq R &= \frac{12\k(\k-1)m\ell^2+5\k q_e^2+24\sqrt{\k}q_b+6\k q_e^2\log(r)}{8(\k-1)\sqrt{\k}\,\ell^2q_b} + \cO(r^2).
\end{align}
We see that these solutions are singular at the origin.

\bibliographystyle{utphys}
\bibliography{ref}

\end{document}